# Emergence of preferred subnetwork for correlated transport in spatial networks: On the ubiquity of force chains in dense disordered granular materials

K. P. Krishnaraj

*Department of Chemical Engineering, Indian Institute of Science, Bangalore, India*
e-mail: krishnarajkp@gmail.com

Abstract

Stress in dense granular materials and other athermal particle aggregates is transmitted through a visually striking subnetwork of interparticle contacts, the filamentary segments of which are referred to as force chains. The emergence of such preferred subnetwork in structurally disordered media with constituents interacting primarily by physical contact is not fully understood. In this work, we study locally correlated transport in Random Geometric Graphs (RGGs), and show the spontaneous emergence of preferred subnetwork. Our findings reveal that, despite structural disorder, system spanning localization of fluxes transmitted through a spatial network can emerge from short ranged correlations. The spatial and statistical features of the subnetwork are surprisingly similar to the strong force network in simulated grain assemblies, and provides insights on the structure and spatial scale of significance of the force chains.

Granular materials, foams, emulsions, gels, suspensions and even biological cells encountered in our daily lives are structurally disordered. Despite their structural disorder, experiments and simulations show that, mechanical stress is transmitted through a visually striking and seemingly ordered subnetwork of interparticle contacts [1-15]. In the past two decades, the spatial and statistical features of the subnetwork have been extensively studied using standard methods [3,5,7,16-18], and recently with ideas inspired from diverse fields like percolation theory [6,19,20], complex networks [20-23] and persistent homology [24,25]. However, the emergence of such preferred subnetwork and its spatial scale of significance has remained elusive. This study aims to address the above questions and are discussed in brief detail next.

The subnetwork of interparticle contacts with normal force, $F_n \geq \langle F_n \rangle$ is known as the strong force network [3], where $\langle F_n \rangle$ is the average normal force over all contacts. The filamentary segments of the strong force network are referred to as force chains. Force chains are system spanning, quasilinear in structure and preferentially oriented in the principal direction of compression [3,5]. Intriguingly, many recent studies have shown that, the force chains are structurally random at large length scales [10,11,19]. Also, scalar pair force correlation measured in experiments on isotropically compressed packings of disks [5] and uniaxially compressed packings of spheres are short ranged [26]. Hence, the interesting question here is, how does a quasilinear yet random preferred subnetwork emerge in a structurally disordered media?

The quasilinear spatial organisation of the strong force network can be quantified using the recently proposed linearity percolation transition [20]. Here, the dependence of the magnitude of force transmitted through a chain of contacts on its linear connectivity $r$ is studied. And, $r$ is defined as the minimum value of $\boldsymbol{n}_i \cdot \boldsymbol{n}_j$ along a chain of contacts, where $\boldsymbol{n}_i$ and $\boldsymbol{n}_j$ are the normal contact vectors of adjacent contacts $i$ and $j$ respectively [20]. The quasilinear spatial organisation is reflected in the variation of the average normal force transmitted ($\overline{F}_n(r)$) through a chain of contact on $r$, and a clear maximum is observed about a critical value of linear connectivity $r_c$. Interestingly, at $r_c$, the scalar pair force correlation is long ranged i.e., exhibits a power law decay [20].

In this study, we model the particle packing as a spatial network with particles as nodes and physical contact between particles as edges [27] and show that, a heterogenous flux transmitting subnetwork spontaneously emerges from locally correlated transport. We reveal that, even in randomly generated graph topologies, short ranged correlations can lead to the spatial localization of fluxes transmitted between nodes. In simulated granular packings, the spatial and statistical features of the subnetwork obtained from transport by persistent self-avoiding walks are surprisingly similar to the strong force network. Our findings provide a unified explanation of various forms of spatial correlation of contact forces reported by previous studies, and we discuss the large-scale structure of force chains and their spatial scale of significance.

Inspired by previous findings on the short-ranged structure of force chains and the strong dependence of the magnitude of force transmitted through a chain on its local structure [7,10,11,20], here we study transport in spatial networks by a locally correlated process. A simple model to study locally correlated transport in spatial networks is the correlated random walk (CRW), widely used in ecology to model animal paths [28]. But, in paths traced by a random walk, loops and self-interactions are possible, hence we study self-avoiding walks (SAWs) with structural correlation between successive step orientations as the model for transport between nodes. We measure the importance of an edge by estimating the fraction of self-avoiding walks that pass through it. If $N_P(i)$ is the number of SAWs that pass-through edge $i$, and $N_P$ is the total number of SAWs sampled, the importance of the edge $i$ is defined as,

$$C_i = \frac{N_P(i)}{N_P} \quad (1)$$

In the self-avoiding walk (SAW) model for transport between nodes, $C$ is a direct measure of flux transmitted through an edge. In social and complex network analysis, such measures are referred to as centrality [29], and is useful in finding the relative influence of a node or edge on information flow through the network. Different measures of centrality have been proposed based on the dynamics of the process of interest or topology of the network [29-31], and a simple random walk betweenness centrality is well-known and widely used in network analysis [29,31]. Also, shortest path betweenness centrality is used to study heat transfer and topology of the contact network in granular materials [22,32], and recently, to predict forces and forecast failure locations in particle packings and disordered lattices respectively [33,34].

To demonstrate the importance of the degree of local structural correlations on transport in a spatial network, we sample paths with increasing structural constraints, from unconstrained to locally correlated SAWs, as described below. The simplest possible path through a network is that traced by a random walk (RW), $P_{RW}$. When a $P_{RW}$ is geometrically constrained such that a constituent node cannot be chosen again, the path obtained is that traced by a SAW, $P_{SAW}$. In $P_{SAW}$, we introduce structural correlation between successive step orientations, based on the unit normals $\boldsymbol{n}_i$ and $\boldsymbol{n}_j$ of adjacent contacts $i$ and $j$ respectively. For a $P_{SAW}$ with local structural correlation of degree $m$, $P_{SAW}^m$, the next contact $j$ with unit normal $\boldsymbol{n}_j$ is chosen with probability,

$$Prob(\boldsymbol{n}_j) = \frac{(\boldsymbol{n}_i \cdot \boldsymbol{n}_j)^m}{\sum_{j=1}^{N_c}(\boldsymbol{n}_i \cdot \boldsymbol{n}_j)^m} \quad (2)$$

Where, $\boldsymbol{n}_i$ is the unit normal of the current contact and $N_c$ is the total number of possible future contacts, and only contacts with $\boldsymbol{n}_i \cdot \boldsymbol{n}_j > 0$ are considered. When $m = 0$, $P_{SAW}^0$ is a locally constrained SAW, and for $m \gg 0$, $P_{SAW}^m$ is locally ordered with large probability.

In RGGs and simulated granular packings, we trace out $P_{SAW}^m$ for varying degree of order $m$ from a randomly chosen edge and direction (every edge has two directions). A SAW is considered complete if any of the following conditions is met,

(i) A boundary node is reached or bulk node with unit coordination number is reached.

(ii) All nodes in contact with the current node do not satisfy the structural constraint or they are constituents of the current path.

Note that, in our definition, for a $P$ to be considered valid minimum two contacts are necessary, and the centrality $C$ estimated based on $P_{SAW}$ are referred to as $C_{SAW}$. For SAWs with degree of local structural order $m$, the centrality is referred to as $C_{SAW}^m$. For every realization, we sample $N_P = 10^6$ paths, and $C$ is estimated for every contact. Our estimates of the spatial organisation and distribution of $C$ does not change with increase in $N_P$ (see Sec. II.D of Supplemental Material (SM) [35]).

We start by studying $2d$ Random Geometric Graphs (RGGs) [36,37]. Our objective of studying RGGs is to show that, a preferred subnetwork can emerge even in randomly generated spatial network, and hence, graph topologies created under mechanical equilibrium or any spatial organisation induced by the constraints of force and torque balances are not a prerequisite. To generate RGGs with coordination number distribution $P(k)$ similar to DEM generated grain assemblies, we use a simple procedure. We random sequentially deposit $N_{circles}$ non-overlapping circles of diameter $d_c$ in a $2d$ $L_x \times L_y$ ($100 d_c \times 100 d_c$) square domain such that the initial area fraction occupied is $\phi \approx 0.50$. And, $d_c$ of every particle is scaled by a factor $\alpha$ such that the required value of $\langle k \rangle$ is achieved (see Sec. II.A of SM for more details [35]).

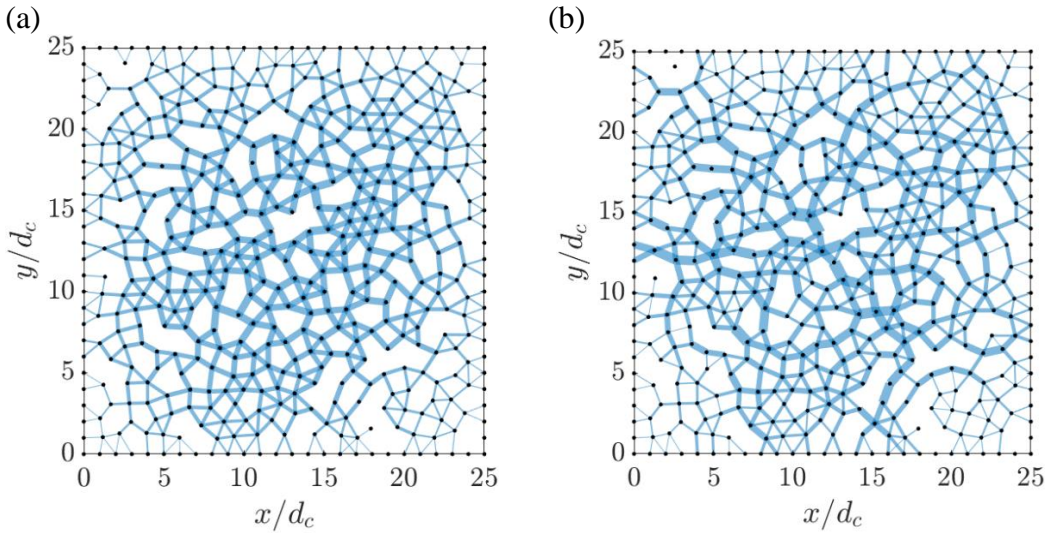

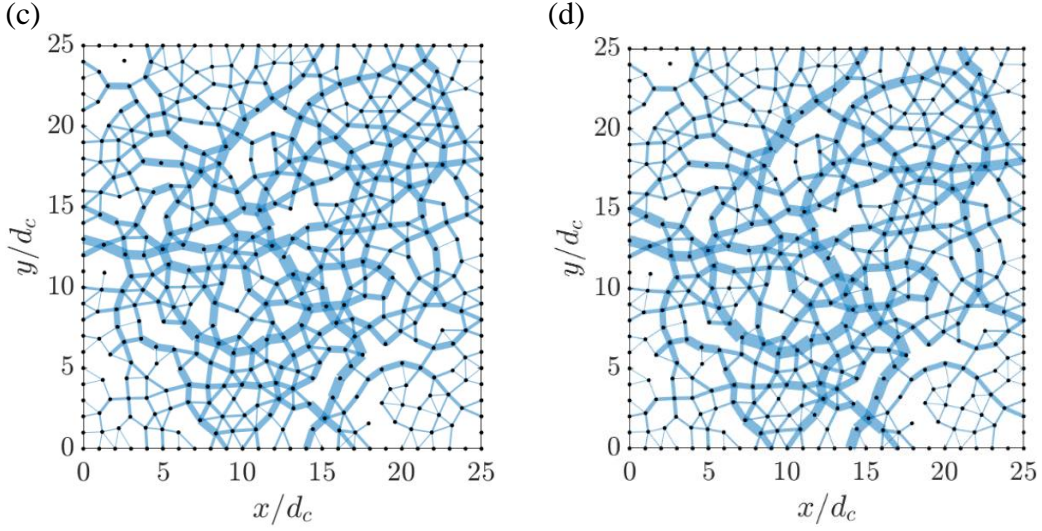

FIG. 1. Spatial organisation of $C$ estimated from simple and correlated SAWs in a randomly chosen realization of RGG with dimensions $L_x \times L_y$ ($25d_c \times 25d_c$) and $\langle k \rangle = 4$. (a) $C_{SAW}$, (b) $C^1_{SAW}$, (c) $C^2_{SAW}$ and (d) $C^4_{SAW}$. The dots represent particle centres and the lines represent edges. The thickness of edges shown in (a-d) are proportional to the magnitude of $C$.

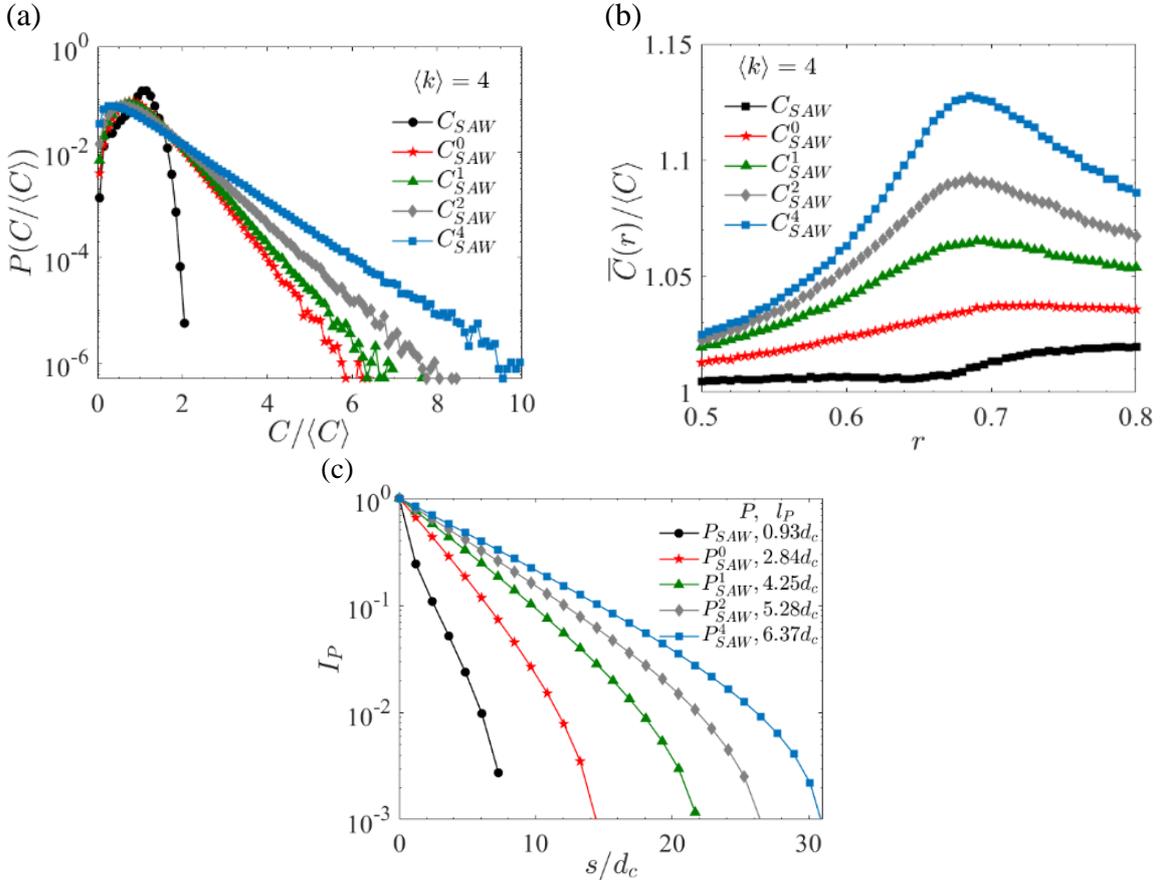

FIG. 2. Statistical features of $C$ estimated from simple and constrained SAWs in RGGs with dimensions $L_x \times L_y$ ($100d_c \times 100d_c$) and $\langle k \rangle = 4$. (a) Probability distribution of $C$, shown in log-linear plot.

(b) Variation of the average value of $C$ in subnetworks with linearity $r$. (c) Persistence of SAWs, shown in log-linear plot (see Sec. II.C of SM for additional information on $l_P$ [35]).

To demonstrate the statistical features of $C$ for various types of paths discussed earlier, we study $C_{SAW}$ and $C_{SAW}^m$ for different degrees of structural order $m$ in RGGs with $\langle k \rangle = 4$. As expected for an unconstrained SAW, the spatial distribution of $C_{SAW}$ is homogenous (Fig. 1(a)). Hence, the probability distribution of $C_{SAW}$ ($P(C_{SAW})$) is sharply peaked about the mean ($\langle C_{SAW} \rangle$), and large fluctuations of $C_{SAW}$ ($C_{SAW} \geq \langle C_{SAW} \rangle$) are negligible (see Fig. 2(a)). However, for $P_{SAW}^4$, where the path is expected to be locally ordered with large probability (Fig. 1(d)), we find $P(C_{SAW}^4)$ to be long tailed with large fluctuations $\approx 10 \langle C_{SAW}^4 \rangle$ (Fig. 2(a)). Now considering $P(C)$ of all types of paths, it is clear that, with increase in local structural order of paths, large fluctuations in $C$ are observed (Fig. 2(a)).

To quantify the spatial organisation of $C$, we study linearity percolation transition in RGGs. Here, subnetworks are identified based on their linear connectivity $r$, we randomly chose a seed contact and find the all the contacts reachable from it for a given value of $r$. We refer to the number of contacts reachable from a seed contact (inclusive) as a subnetwork. The statistics reported are averaged over a large number of subnetworks (see Sec. I.D of SM for more details [35]). In subnetworks with linear connectivity $r$, $\overline{C}(r)$ shows a clear maximum about $\sim r_c$ i.e., large $C$ segments are quasilinear in structure (Fig. 2(b)). And, the maximum value of $\overline{C}(r)$ increases with the local structural order of the paths sampled (Fig. 2(b)), clearly shows the emergence of preferred subnetwork (Fig. 1(b-d)). Our estimate of $r_c$ for RGGs with $\langle k \rangle = 4$ is 0.6279, see Sec. II.B of SM [35].

A simple measure of characteristic distance of structural correlation along a path $P$ is the persistence, $l_P$ [10,11,38]. Here, $l_P = \langle n_1 . n_H \rangle$ is the measure of persistence of seed contact direction $n_1$, and $n_H$ is the direction of the contact at step $H$ along the contour of the path, the angle brackets indicate averaging over paths with at least $H$ steps. And, $s(H)$ is the average length along the contour of paths with at least $H$ steps, see Sec. II.C of SM for more details [35]. As shown in Fig. 2(c), the length of directional persistence of the paths $l_P$ increases with the structural order $m$, and the corresponding large $C$ subnetworks are increasingly linear in structure (Fig. 1(b-d)). Importantly, $l_P \approx 6.4 d_c$ or short ranged even for SAWs with strong local correlation ($m = 4$), and this clearly shows that, large-scale preferred subnetwork can emerge from locally or short-ranged correlated transport in structurally disordered spatial networks. Next, we study $C$ in simulated grain assemblies.

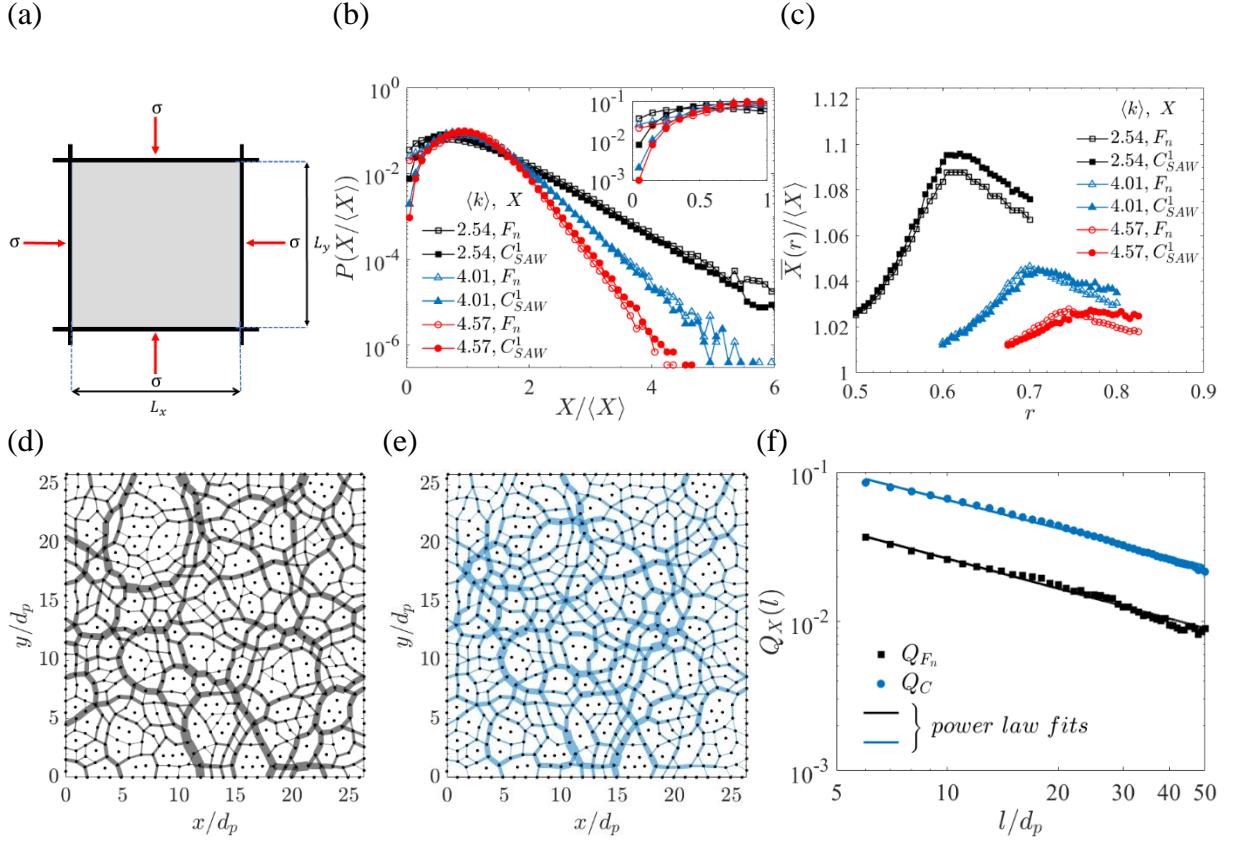

FIG. 3. Spatial and statistical features of $F_n$ and $C_{SAW}^1$ in simulated $2d$ isotropically compressed packings. The dimensions of packings studied in plots (a-c and f) are $L_x \times L_y$ ($100 d_p \times 100 d_p$). (a) Schematic of $2d$ isotropic compression, $\sigma$ is the stress applied at the boundary. (b) Probability distribution of $X$ ($X = F_n$ or $C_{SAW}^1$), the inset shows $P(X)$ for $X \leq \langle X \rangle$, and both are shown in log-linear plot. (c) Variation of the average value of $F_n$ and $C_{SAW}^1$ in subnetworks with linearity $r$. (d, e) Spatial organisation of $F_n$ (d) and $C_{SAW}^1$ (e) in a randomly chosen realization with dimensions $L_x \times L_y \sim (25 d_p \times 25 d_p)$ and $\langle k \rangle = 2.54$. The dots represent particle centres and the lines represent edges. The thickness of edges shown in (d, e) are proportional to the magnitude of $F_n$ or $C_{SAW}^1$. (f) $Q_{F_n}$ and $Q_C$ in packings with $\langle k \rangle = 2.71$ at critical linearity $r_c = 0.609$ [20], shown in log-log plot. The coefficients of power law fits ($a \ell^b$) shown are $a = 0.12, b = -0.67$ and $a = 0.28, b = -0.63$ for $F_n$ and $C_{SAW}^1$ respectively.

To make quantitative comparison with the contact network in granular materials, we first study $C$ in simulated $2d$ isotropically compressed packings, with mean particle diameter $d_p$ in a $L_x \times L_y$ ($100 d_p \times 100 d_p$) square domain (see Fig. 3(a)), using the Discrete Element Method (DEM) [39] and performed with the open-source molecular dynamics package LAMMPS [40], additional details of the simulation method and creation of granular packings are provided in Sec. I of SM [35]. We find that, the spatial and statistical features of $C$ estimated from self-avoiding paths with linear local structural order ($P_{SAW}^1$) are remarkably similar to the strong force network in granular materials. In isotropically compressed packings, $P(C_{SAW}^1)$ for $C_{SAW}^1 \geq \langle C_{SAW}^1 \rangle$ is in close agreement with $P(F_n)$ for $F_n \geq \langle F_n \rangle$ (Fig. 3(b)). The deviation of the form of $P(C_{SAW}^1)$ as $C_{SAW}^1 \to 0$ from $P(F_n)$ as $F_n \to 0$ suggests that weak forces in granular materials are not spatially correlated (see inset in Fig. 3(b)) or some features of force transmission are not adequately represented by the persistent SAW model. As shown in Fig. 3(d, e), large $C$ subnetworks are quasilinear in structure, and $\overline{C_{SAW}^1}(r)$ is in good agreement with $\overline{F}_n(r)$ as $\sim r_c$ is approached (Fig. 3(c)).

With increase in $\langle k \rangle$ or $\phi$, the magnitude of fluctuations of $P(F_n)$ for $F_n \geq \langle F_n \rangle$ and $P(C_{SAW}^1)$ for $C_{SAW}^1 \geq \langle C_{SAW}^1 \rangle$ decreases (Fig. 3(b)), and hints a spatially homogenous distribution of $F_n$, also reported by previous studies [41-43]. The spatial homogeneity is quantitatively shown by the decrease in the maximum value of $\overline{F}_n(r)$ and $\overline{C_{SAW}^1}(r)$ about $r_c$ (Fig. 3(c)). Interestingly, the persistence length $l_P$ of $P_{SAW}^1$ is almost the same in packings with varied $\langle k \rangle$ or $\phi$ (see Sec. II.C of SM [35]), hence in addition to the degree of local structural correlation, the topology of the contact network strongly influences correlated transport in particle packings. Next, we discuss the spatial correlation of $F_n$ reported by previous studies in detail.

The spatial scale of correlation of the contact forces in granular materials and other forms of athermal disordered media has been the focus of many recent studies [5,6,10,11,19,20]. Studies on the structure of force chains and force percolation transition suggest that, the subnetworks of large force transmitting contacts are structurally random at large length scales [6,10,11,19]. And, recent study based on linearity percolation shows the spatial correlation of contact force magnitudes in subnetworks at $r_c$ to be long ranged [20], in contrast to short ranged correlation for the complete network [5]. Here, we find that, the insights on spatial correlation reported by previous studies are signatures of locally correlated transport. To demonstrate, we define the spatial correlation function as [20],

$$Q_X(l) = \langle \delta(l_{ij} - l) X_i' X_j' \rangle \tag{3}$$

where $X_i'$ is the deviation of $F_n$ or $C$ at contact $i$ from $\langle F_n \rangle$ or $\langle C \rangle$ respectively, $l_{ij}$ is the distance between contacts $i$ and $j$, and the angle brackets denote averaging over the subnetwork of linearity $r$ over multiple configurations. Here, in subnetworks of granular packings at $r_c$, we find the spatial correlation of centrality $Q_C(l)$ to be long ranged and exhibit power law decay similar to $Q_{F_n}(l)$, and is shown in Fig. 3(f). However, the persistence length $l_P$ of the $P_{SAW}^1$ is short ranged $\approx 4d_p$ (see Sec. II.C of SM [35]), revealing the loss of structural order at large length scales. In other words, the intriguing long ranged correlation of contact forces observed at $r_c$ is also a signature of transport along locally ordered paths in granular materials.

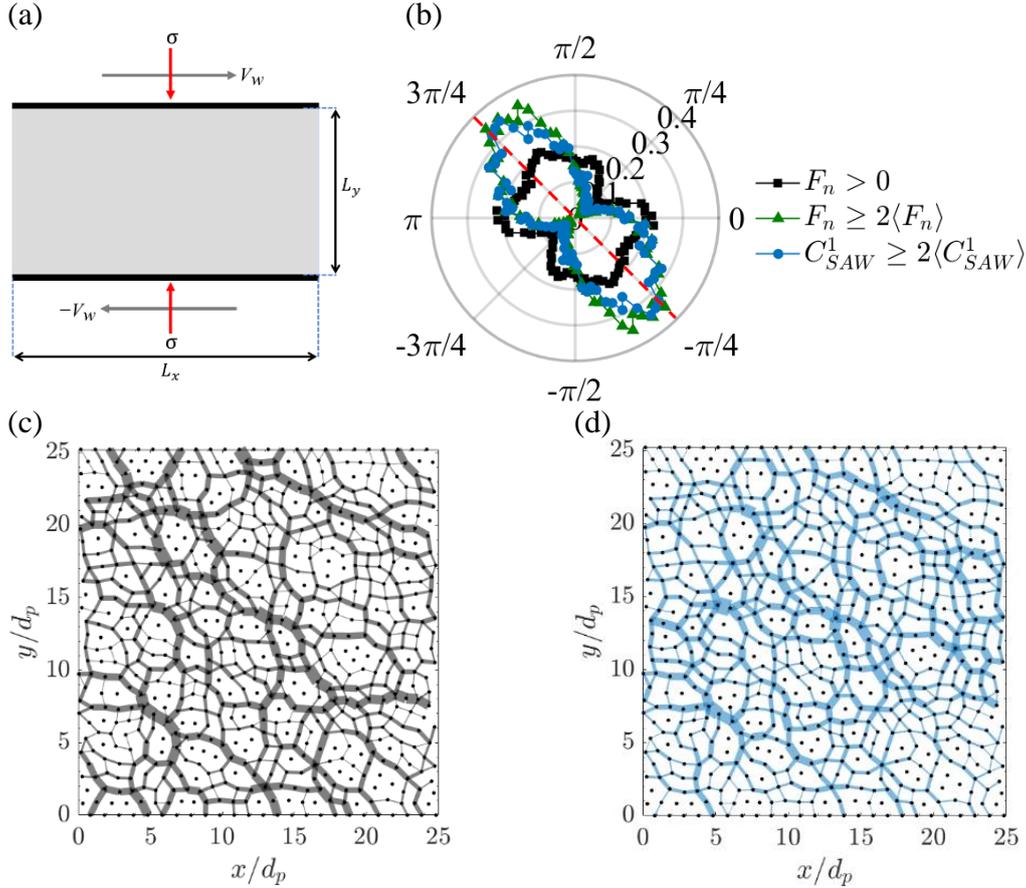

FIG. 4. Spatial and statistical features of $F_n$ and $C_{SAW}^1$ in simulated $2d$ sheared packings with dimensions $L_x \times L_y$ $(100d_p \times 100d_p)$, $\langle k \rangle = 2.73$ and Savage number $Sa = 10^{-8}$. (a) Schematic of plane shear in $2d$, $\sigma$ is the applied stress and $V_w$ is the velocity of the confining walls. (b) Probability distribution of contact vector orientations, $P(\theta)$, the red dashed line indicates the principal direction of compression. (c, d) Spatial organisation of $F_n$ (c) and $C_{SAW}^1$ (d) in a randomly chosen realization with dimensions $L_x \times L_y$ $(25d_p \times 25d_p)$ and $\langle k \rangle = 2.73$. The dots represent particle centres and the lines represent edges. The thickness of edges shown in (c, d) are proportional to the magnitude of $F_n$ or $C_{SAW}^1$.

Finally, we study $C_{SAW}^1$ in anisotropic granular packings created by plane shear (Fig. 4(a)) (see Sec. I.B of SM for more details [35]). Here, in addition to the spatial organisation (see Sec. I.C of SM [35]), we find that the contacts in large $C_{SAW}^1$ subnetworks ($C_{SAW}^1 \geq \langle C_{SAW}^1 \rangle$) are preferentially oriented in the principal direction of compression, and is remarkably similar to the strong force network ($F_n \geq \langle F_n \rangle$), see Fig. 4(b-d).

In conclusion, we have shown that, short ranged correlations can lead to localization of fluxes transmitted between nodes in a spatial network. Consequently, a system spanning preferred subnetwork emerges even in structurally disordered networks. Our findings in RGGs reveal that, such preferred subnetwork can emerge in randomly generated spatial networks, and hence, graph topologies created under mechanical equilibrium or any spatial organisation induced by the constraints of force and torque balances on particles are not a prerequisite. The strong similarities between the spatial and statistical features of the subnetworks obtained from persistent SAWs and the strong force network suggests that a locally correlated process as the plausible mechanism for force transmission in granular packings. As many recent experiments have uncovered striking visual and statistical resemblance in the manner stress is transmitted in granular materials, foams, emulsions, gels, suspensions and even biological cells [1-15], our results could potentially explain the ubiquity of force chains in athermal disordered media. Furthermore, given the information of the structure of the packing, our study opens up an interesting possibility of modelling force transmission in granular materials as flows in spatial networks. Such an approach to understand classic and puzzling problems in the mechanical behaviour of granular materials will be presented elsewhere [44].

## Acknowledgements

I am profoundly grateful to N.Marayee, P.Rajamani, K.P.Dhanasekaran, A.K.K.Arvind, Mani, R.S.Veeraraahavan and R.Karthick Sundar for their encouragement and support.## References

# Supplemental material

**Supplemental note I**
**A. Particle dynamics simulations**

We use the Discrete Element Method widely used for simulation of granular materials [39]. The simulations described in this study were conducted using the open-source molecular dynamics package LAMMPS [40]. To simulate dissipative interaction between granular particles, we use the standard linear spring and dashpot model. In addition, a Coulomb slider is used to incorporate rate-independent frictional force in the linear spring and dashpot modules in the tangential direction of the contact (Fig. S1). A complete description of the contact model and DEM implementation used in this study is provided in Ref. [20].

The parameters used in the simulations are listed in Fig. S1. The values of $k_n$, $k_t$ and $\gamma_t$ was chosen based on previous studies [20, R1] that have attempted to model hard grains such as glass beads and sand. The value of $\gamma_n$ chosen is such that the normal coefficient of restitution $e_n$ is 0.7. In all our computations, $\mu$ is set to 0.5. The $2d$ simulations were conducted by placing spheres in a plane and allowing movement only within the plane.

The particle sizes were chosen from a uniform distribution with lower and upper limits of $0.8d_p$ and $1.2d_p$ respectively, where $d_p$ is the mean diameter. The walls were constructed with particles of diameter $d_p$ set in a close packed linear lattice in $2d$. In all the simulations, the constants characterizing grain-wall interactions are the same as those for grain-grain interactions.

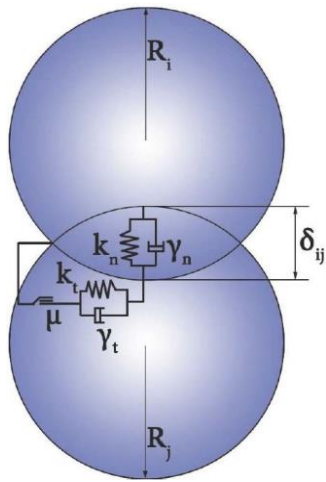

| Parameter | Value |
|---|---|
| $k_n$ | $10^5$ |
| $k_t$ | $\frac{2}{7} k_n$ |
| $\gamma_n$ | 28571 |
| $\gamma_t$ | $\frac{1}{2} \gamma_n$ |

FIG. S1. Schematic of the soft particle interaction model between spheres of radii $R_i$ and $R_j$. The values of parameters used in the model are given in the table.

## B. Creation of granular packings

We use the protocol detailed in Ref. [20] for creating the $2d$ isotropically compressed and sheared granular packings. The stress applied at the boundaries $\sigma = F/Ld_p$, where $F$ is the force applied on the boundary, $L$ is the size of the system and $d_p$ is the mean diameter of the particle ($L = L_x = L_y$ for isotropic compression). For isotropic packings shown in Fig. 3, stress applied on the boundaries $\sigma$, and the corresponding average area fraction $\phi$ and coordination number $\langle k \rangle$ are given in Table S1. The unit of $\sigma$ provided in Table S1 is $kg\ m^{-1}s^{-2}$.

The Savage number $Sa$ [R2] for the plane shear study is $\approx 10^{-8}$. Hence, the granular material is in the quasistatic flow regime. And, $\sigma = 100\ kg\ m^{-1}s^{-2}$, $\phi = 0.8133$ and $\langle k \rangle = 2.7346$ for sheared granular assemblies studied in Fig. 4.

The results of isotropically compressed and sheared granular assemblies reported in this study are averages over 100 independent realizations.

| Preparation method | $\sigma$ | $\phi$ | $\langle k \rangle$ |
|---|---|---|---|
| Isotropic compression | 0.36 | 0.8101 | 2.5353 |
| Isotropic compression | 325.0 | 0.8132 | 2.7147 |
| Isotropic compression | $1.67 \times 10^5$ | 0.8538 | 4.0109 |
| Isotropic compression | $5.18 \times 10^5$ | 0.9138 | 4.5670 |
| Plane shear | 100 | 0.8133 | 2.7346 |

Table S1. Types of $2d$ granular packings studied, and $\phi$ and $\langle k \rangle$ corresponding to $\sigma$ applied at the boundary are listed.

## C. Spatial structure and probability distribution of $C$ in $2d$ sheared granular packings

In $2d$ sheared grain assemblies shown in Fig. 4, the functional form of $P(C_{SAW}^1)$ for $C_{SAW}^1 \geq \langle C_{SAW}^1 \rangle$ is similar to $P(F_n)$ with $F_n \geq \langle F_n \rangle$ (Fig. S2(a)). And, the spatial structure of $\overline{C_{SAW}^1}(r)$ is in close agreement with $\overline{F}_n(r)$ as $r_c$ is approached, shown in Fig. S2(b).

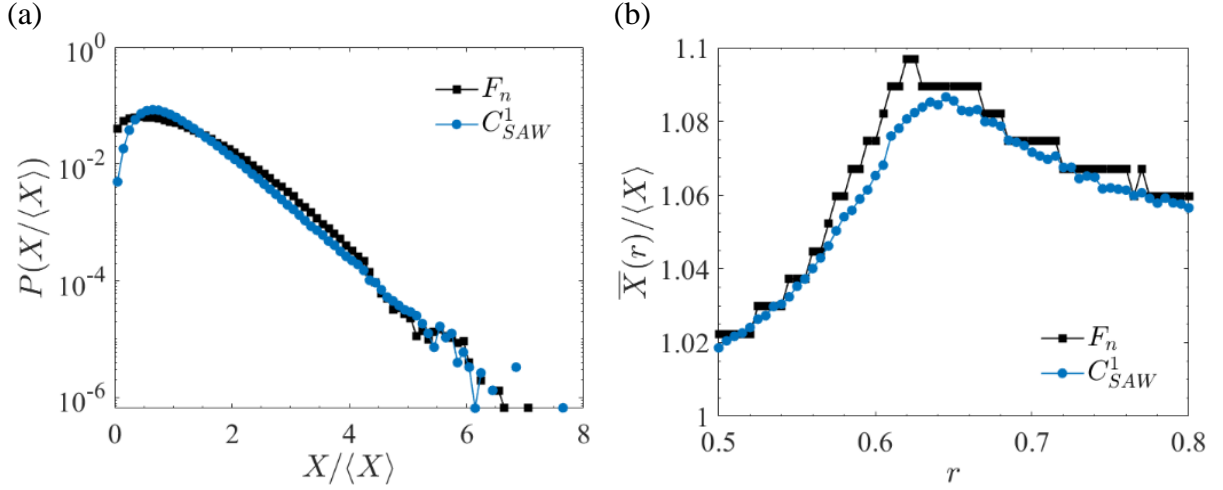

FIG. S2. Spatial and statistical features of $F_n$ and $C_{SAW}^1$ in simulated $2d$ sheared packings with dimensions $L_x \times L_y$ $(100d_p \times 100d_p)$, $\langle k \rangle = 2.73$ and Savage number $Sa = 10^{-8}$. (a) Probability distribution of $X$ ($X = F_n$ or $C_{SAW}^1$). (b) Variation of the average value of $F_n$ and $C_{SAW}^1$ in subnetworks with linearity $r$.

**D. Linearity percolation transition in RGGs and simulated grain assemblies**

In RGGs and simulated grain assemblies, we find the subnetwork corresponding to a contact for the given value of linearity $r$ using the procedure described in Ref. [20]. Here, $r$ is defined as the minimum value of $\mathbf{n}_i \cdot \mathbf{n}_j$ along a path $P$, where $\mathbf{n}_i$ and $\mathbf{n}_j$ are the normal contact vectors of adjacent contacts $i$ and $j$ respectively. We transform the particle packing to a directed graph and use the Breadth First Search algorithm as detailed in Ref. [20]. In RGGs and simulated grain assemblies, we randomly chose a seed contact and find the all the contacts reachable from it for a given value of $r$. We refer to the number of contacts reachable from a seed contact (inclusive) as a subnetwork. For every independent realization, the statistics of linearity percolation reported in this study are averaged over a large number of subnetworks (10% of the total number of contacts), as discussed in Ref. [20].

An interesting observation was made in Ref. [20], there exists a critical value of linearity $r_c$ at which the spatial correlation function ($Q_F(l)$ defined in Eq. (3)) is long ranged i.e., exhibits a power law decay with distance. Paths with linearity $\sim r_c$ are system spanning and exhibit local structural order, and contain the force chains [20]. In simple terms, paths with linearity $r_c$ can be considered as system spanning yet maximally linear in a given spatial network or particle packing.

**Supplemental note II**

**A. Creation of Random Geometric Graphs (RGGs)**

We randomly place $N_{circles}$ non-overlapping circles of diameter $d_c$ in a $2d$ square domain with spatial dimensions $100d_c \times 100d_c$ ($L_x \times L_y$). The initial spatial configuration of the circles was created by random sequential deposition i.e., a new circle placed in a randomly chosen coordinate is accepted only if does not overlap with existing circles, and this process is repeated until the required number of circles are created. The initial area fraction occupied by the circles $\phi \approx 0.50$, and hence excluding boundary particles, $N_{circles} = 6363$. To achieve the desired probability distribution of coordination number $k$ ($P(k)$) similar to DEM generated grain assemblies, we increase $d_c$ of every particle by a scale factor α. As shown in Fig. S3(a), with increase in the value of α, the average coordination number $\langle k \rangle$ of the graph increases. The coordination number distribution of RGGs obtained from the above procedure is similar to DEM generated $2d$ isotropic particle packings, see Fig. S3(b, c). The results shown in Fig. 1 and 2 for RGGs are averages over 300 independent realizations.

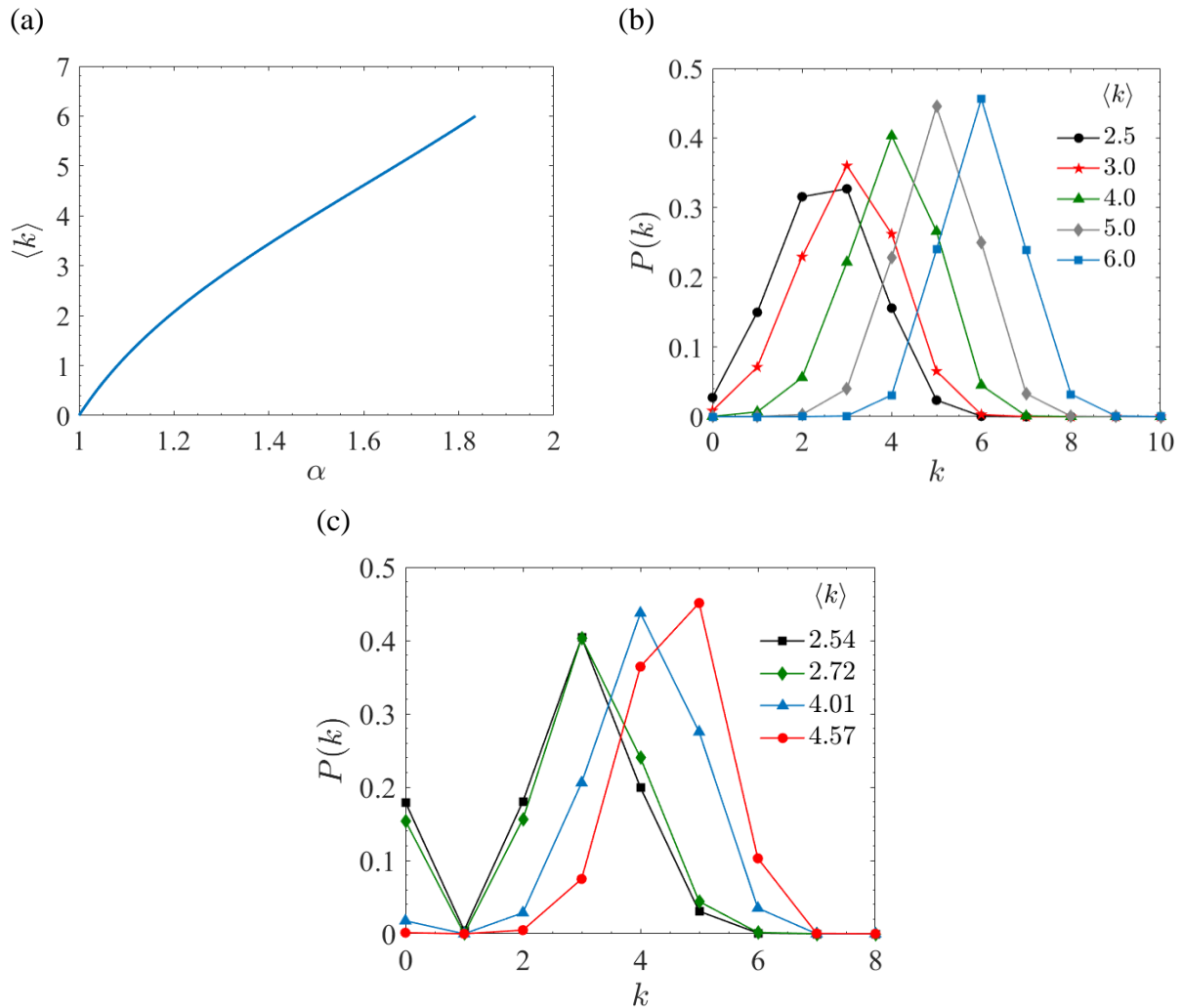

FIG. S3. Creation of RGGs. (a) The variation of average coordination number $\langle k \rangle$ with increase in scale factor $\alpha$ in RGGs, and was averaged over 300 independent realizations. (b, c) Coordination number distribution $P(k)$ in RGGs (b) and in isotropically compressed granular packings (c), both are averages over 100 independent realizations.

## B. Critical value of linearity of RGGs

We estimate the critical value of linearity of RGGs generated from the procedure described in Sec. II.A using the standard method described in Ref. [R3]. As shown in Fig. S4, we find a good collapse of the data corresponding to different system sizes when $P(r)$ is plotted against $(r - r_c) N^{\frac{1}{d\nu}}$, where $r_c$ is the critical value of linearity, $d$ is the spatial dimension and $\nu$ is the correlation length exponent. Our estimate of $r_c$ for RGGs with $\langle k \rangle = 4$ is 0.6279, and $\nu = 1.3839$.

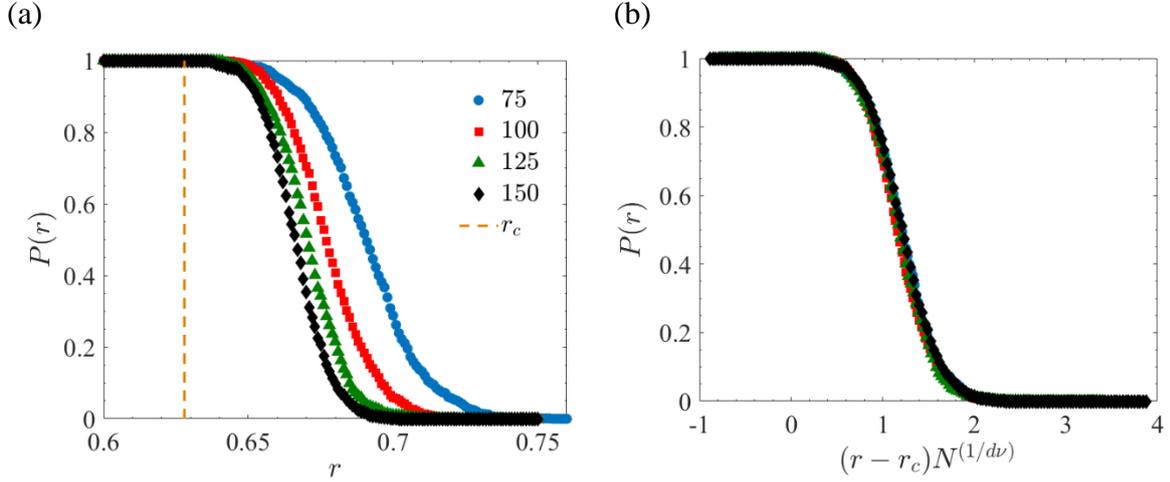

FIG. S4. Critical value of linearity in RGGs with $\langle k \rangle = 4$. (a) Percolation probability for different system sizes; the system size $L$ is in units of initial diameter of the circles $d_c$. The dashed line shows $r_c$. (b) Collapse of all the data when $P(r)$ is plotted against $(r - r_c) N^{\frac{1}{d\nu}}$. Here, $d$ is the spatial dimension. $P(r)$ shown in (a) and (b) are averages over 300 independent realizations.

## C. Contour distance $s$ and persistence length $l_P$ of SAWs

The length of directional persistence or persistence length of SAWs $l_P$ is found by fitting the variation of persistence $I_P$ with distance $s$ to an exponential function of the form, $\beta e^{\left(\frac{-s}{l_P}\right)}$ [38], where $\beta$ is a constant. Here, $s(H)$ is the average length along the contour of paths with at least $H$ steps, and is shown in Fig. S5(a, b). To estimate $l_P$, we sample $10^6$ SAWs in every RGG and DEM generated granular packing, and further averaged over 100 independent realizations. $I_P$ and $l_P$ of $P^1_{SAW}$ in $2d$ isotropically compressed packings are shown in Fig. S5(c).

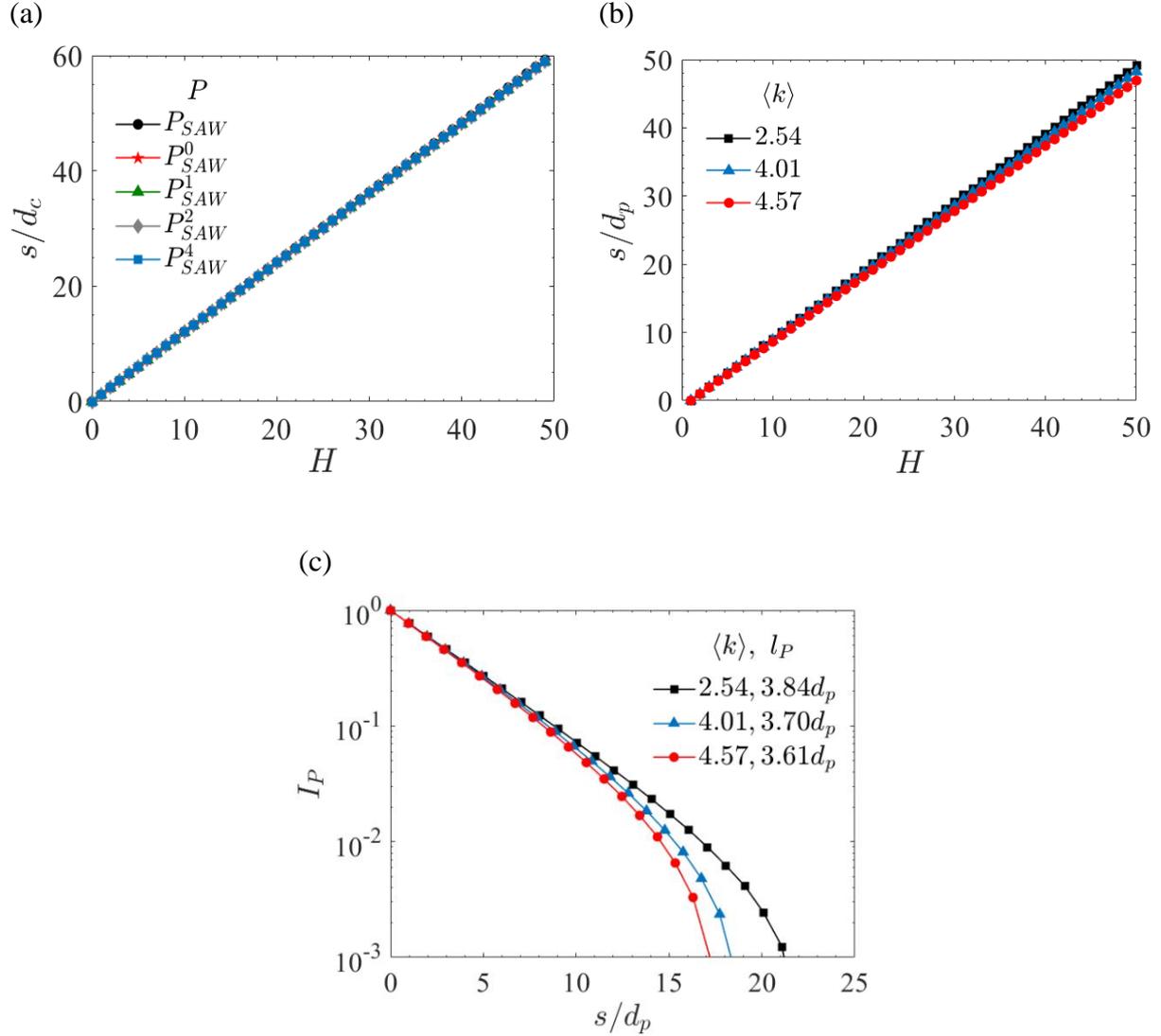

Fig. S5. Contour distance and persistence of SAWs in RGGs and isotropically compressed granular packings. (a, b) Contour distance $s$ of SAWs with at least $H$ steps, (a) RGGs with $\langle k \rangle = 4$, (b) $P^1_{SAW}$ in $2d$ isotropically compressed granular packings and (c) Persistence $I_P$ of $P^1_{SAW}$ in $2d$ isotropically compressed granular packings.

### D. Number of SAWs sampled, $N_P$

In this study, the number of SAWs sampled in RGGs and granular packings are $N_P = 10^6$. With increase in $N_P$, $P(C)$ and $\overline{C}(r)$ does not change considerably, and is shown in Fig. S6 (a-d).

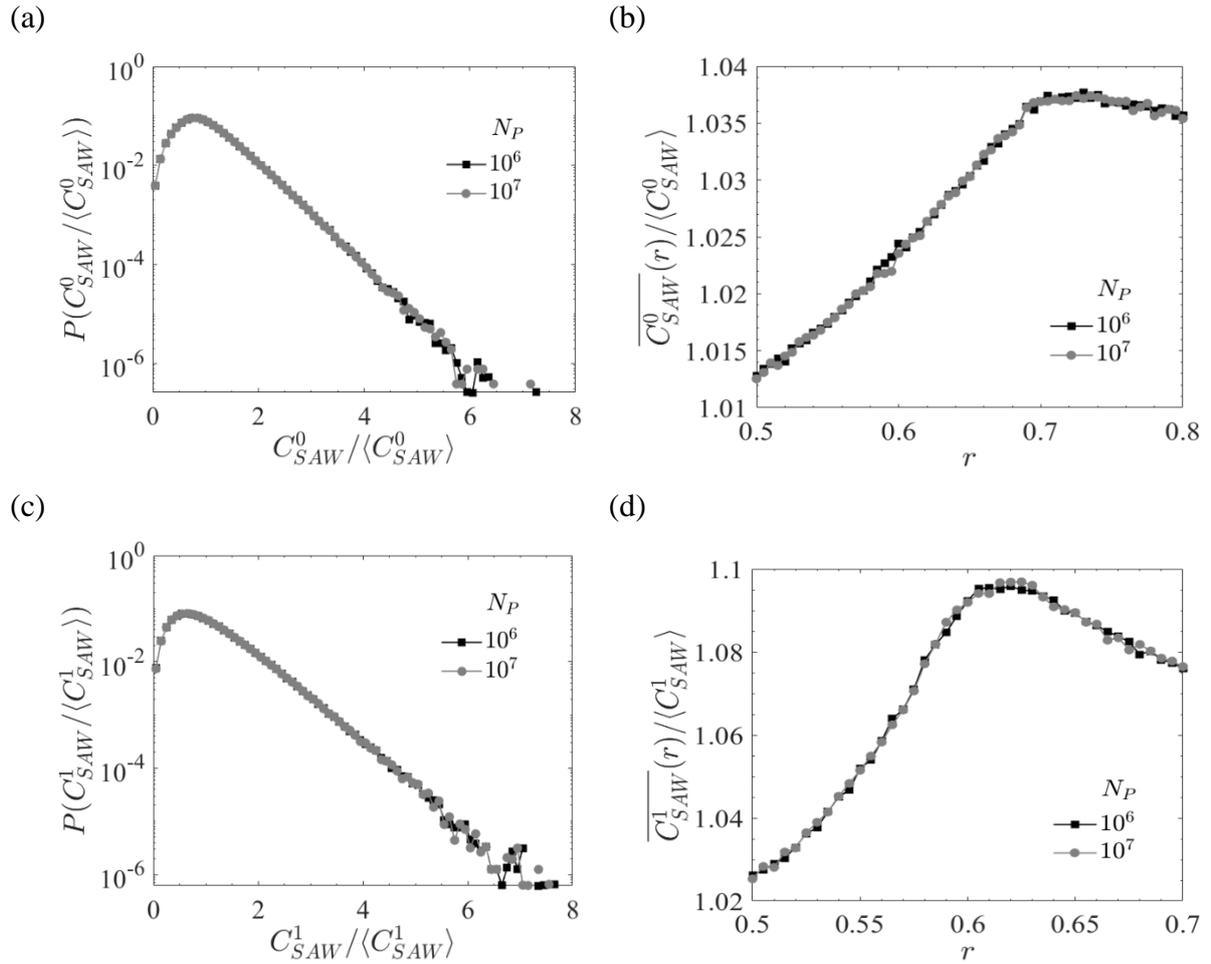

Fig. S6. Significance of the number of self-avoiding paths sampled, $N_P$. (a) Probability distribution of $C$ in RGGs with $\langle k \rangle = 4$, (b) Variation of $\overline{C}(r)$ in subnetworks with linearity $r$ in RGGs with $\langle k \rangle = 4$, (c) Probability distribution of $C$ in isotropically compressed $2d$ granular packings with $\langle k \rangle = 2.54$ and (d) Variation of $\overline{C}(r)$ in subnetworks with linearity $r$ in isotropically compressed $2d$ granular packings with $\langle k \rangle = 2.54$.

**Supplemental references**